\documentclass{article}
\usepackage{icmc2020template}
\usepackage{times}
\usepackage{ifpdf}
\usepackage{mathtools}
\usepackage{soul}
\usepackage[english]{babel}

\def\papertitle{Learning Models for Query by Vocal Percussion: A Comparative Study}

\def\firstauthor{Alejandro Delgado}
\def\secondauthor{SKoT McDonald}
\def\thirdauthor{Ning Xu}
\def\fourthauthor{Charalampos Saitis}
\def\fifthauthor{Mark Sandler}

\newif\ifpdf
\ifx\pdfoutput\relax
\else
\ifcase\pdfoutput
\pdffalse
\else
\pdftrue
\fi
\fi

\ifpdf % compiling with pdflatex
\usepackage[pdftex,
pdftitle={\papertitle},
pdfauthor={\firstauthor, \secondauthor, \thirdauthor},
bookmarksnumbered, % use section numbers with bookmarks
pdfstartview=XYZ % start with zoom=100% instead of full screen; 
]{hyperref}
\usepackage[pdftex]{graphicx}
\graphicspath{{./figures/}}
\DeclareGraphicsExtensions{.pdf,.jpeg,.png}

\usepackage[figure,table]{hypcap}

\else % compiling with latex
\usepackage[dvips,
bookmarksnumbered, % use section numbers with bookmarks
pdfstartview=XYZ % start with zoom=100% instead of full screen
]{hyperref} % hyperrefs are active in the pdf file after conversion

\usepackage[dvips]{epsfig,graphicx}
\graphicspath{{./figures/}}
\DeclareGraphicsExtensions{.eps}

\usepackage[figure,table]{hypcap}
\fi

\hypersetup{
colorlinks,%
citecolor=black,%
filecolor=black,%
linkcolor=black,%
urlcolor=black
}

\title{\papertitle}

\fiveauthors
{0.5in}
{\firstauthor} {Roli / Queen Mary University of London \\ %
{\tt \href{mailto:alejandro@roli.com}{alejandro@roli.com}}}
{\secondauthor} {Roli \\ %
{\tt \href{mailto:skot@roli.com}{skot@roli.com}}}
{\thirdauthor} {Roli \\ %
{\tt \href{mailto:ning@roli.com}{ning@roli.com}}}
{\fourthauthor} {Queen Mary University of London \\ %
{\tt \href{mailto:c.saitis@qmul.ac.uk}{c.saitis@qmul.ac.uk}}}
{\fifthauthor} {Queen Mary University of London \\ %
{\tt \href{mailto:mark.sandler@qmul.ac.uk}{mark.sandler@qmul.ac.uk}}}

\begin{document}
\capstartfalse
\maketitle
\capstarttrue
\begin{abstract}

The imitation of percussive sounds via the human voice is a natural and effective tool for communicating rhythmic ideas on the fly. Thus, the automatic retrieval of drum sounds using vocal percussion can help artists prototype drum patterns in a comfortable and quick way, smoothing the creative workflow as a result. Here we explore different strategies to perform this type of query, making use of both traditional machine learning algorithms and recent deep learning techniques. The main hyperparameters from the models involved are carefully selected by feeding performance metrics to a grid search algorithm. We also look into several audio data augmentation techniques, which can potentially regularise deep learning models and improve generalisation. We compare the final performances in terms of effectiveness (classification accuracy), efficiency (computational speed), stability (performance consistency), and interpretability (decision patterns), and discuss the relevance of these results when it comes to the design of successful query-by-vocal-percussion systems.

\end{abstract}

%The abstract should be placed at the start of the top left column and should contain about 150-200 words. The abstract should be formatted in italic type (this has already been set in the abstract style).

\section{Introduction}
\label{s1}

One of the aims of Music Information Retrieval (MIR) as a field of research is to provide musicians and music producers with new tools that can make their workflow more pleasant. Thanks to this, mechanical and sometimes laborious tasks can now be fully automated, allowing artists to better focus on the most creative aspects of their craft. Subfields of MIR like musical key detection, chord estimation, and automatic mixing are dedicated to the development of these methods.

Another subfield that shares the same goal is Query by Example (QE). It investigates how to quickly query a specific audio file from a sound library by providing another file that sounds like it, like a sketch recording. The searching process is usually fully audio content-based, and therefore does not require any other kind of metadata like labels or text descriptions. A special case of QE is that of Query by Vocal Imitation (QVI), in which the example audio file fed to the system contains a vocal imitation of the desired sound. The subject covered in this study, \textit{Query by Vocal Percussion} (QVP), is a form of QVI in which percussive sounds are retrieved from percussive vocal imitations. In particular, we focus on retrieving four types of drum sounds: kick drum, snare drum, closed hi-hat, and opened hi-hat.

Querying drum sounds by vocal percussion requires distinct approaches given the context and resources available. For instance, a basic ``naive" algorithm could simply retrieve the most relevant drum samples by objective sound content similarity with the vocal imitation (e.g. by spectral distance) \cite{25}. Another more nuanced case-study is when one wants to retrieve the sound of a snare drum from a particular brand or articulated in a specific manner \cite{9}. Here, the algorithms must be aware of both the different sounds that snare drums can make and how differently the user vocalises them. This is, therefore, a problem of correspondence between two sounds, and the main objective is to find the set of audio features that best link them. Lastly, if one wants to trigger a fixed drum sample when a specific vocal imitation is given, then the algorithms are just required to learn how to distinguish between vocal imitations alone, meant to trigger different drum samples \cite{8}. This is a problem of correspondence between a sound and a label (classification), and the relevant set of audio features would now be the one that best separates all classes. The focus of the present work is put on this latter task.

A literature review of relevant past work on QVP is given in section \ref{s2}. The methods and sub-routines to study are laid out in section \ref{s3} and we give a detailed summary of the final results in section \ref{s4}. We conclude with a brief recapitulation and future perspectives in section \ref{s5}.

% irrespective of their underlying nature

% , hit with a certain type of stick or

% A very important thing to note in this type of QVP is that it is not reducible to regular sound event classification task when considering a realistic scenario. Here, we use two datasets of a slightly different nature. For the first dataset, the users of the QVP system must show the algorithm how they imitate a kick drum, for example, by recording some isolated examples for analysis. Then, the algorithm will be able to recognise kick drum utterances in a beatbox improvisation, and the same happens with the rest of the drum types. The challenge on this is that

\section{Problem Definition and Past Work}
\label{s2}

An important thing to note about our approach to performing QVP is that, when considering a realistic scenario, it is not entirely reducible to a regular sound event classification task. Here the users are not supposed to label the utterances they record for the algorithm to train successfully, so the labels have to be made implicit in some way. This is the reason why the final training and testing datasets end up being of a slightly different nature.

For the \textit{training dataset}, the users of the QVP system record several utterances of each drum type in four separated audio files (kick drum, snare drum, closed hi-hat, and opened hi-hat). This allows the algorithm to automatically assign a different class to sounds depending on the file that contains them. In this way, all the utterances are automatically labelled without requiring the users to do it themselves. For the \textit{testing dataset}, the users record a beatbox-style improvisation with utterances of mixed classes in a single file, without label annotations. The trained system would try to correctly classify the vocalisations despite the expressive resources in the improvised performance, like modulations in pitch, loudness, or duration. The challenge is, then, to extract the most descriptive audio features possible in the training process so that the algorithm is robust enough to handle these expressive nuances in the testing data.

Most relevant past work on QVP is concerned with utterance classification rather than querying. This means that only one of the two datasets (isolated utterances \cite{17} or full improvisation \cite{14}) is used for both training and testing, generally resulting in higher classification accuracies\footnote{We obtain 98\% classification accuracy on this task using our dataset}. The works most often feature three classes of imitated sounds (kick drum, snare drum, and hi-hat) instead of four. While automatic onset detection is explored as well in some of these projects, our study is exclusively focused on the classification of vocal percussion utterances into their correct drum types.

In \cite{17}, Hazan and Ramirez feed spectral and temporal audio features from the vocal percussion utterances to a decision tree algorithm, reporting results of up to 90\% classification accuracy in testing sets. Sinyor et al. also explored the AdaBoost algorithm in \cite{16}, achieving the best classification results (98\%) together with a decision tree. Nakano et al. used Mel Frequency Cepstral Coefficients and a Hidden Markov Model to classify drum patterns containing kick and snare drums in \cite{19}, hitting 93\% accuracy. Kapur et al. \cite{4} fed different types of audio low-level descriptors to an artificial neural network and investigated how well each of them could perform the task on its own. They found that the zero-crossing rate descriptor was the most informative, reaching an accuracy of 97\%. Stowell et al. \cite{14} observed that one could obtain higher classification accuracies in beatbox utterance classification by delaying the start of the first analysis frame 23 milliseconds from each event's onset, ignoring the information in the sound's transient. Finally, in \cite{2}, Ramires et al. used a set of engineered features and a K-Nearest Neighbours algorithm to build a user-based vocal percussion classifier that consistently performed better than a general-purpose one. They also introduce a QVP plugin for Digital Audio Workstations in \cite{8}.

We draw inspiration from the studies above when selecting audio descriptors and machine learning algorithms to classify the utterances. To our knowledge, this is the first study that introduces data augmentation techniques and deep learning strategies for QVP.

\section{Methodology}
\label{s3}

\begin{figure}
\label{f1}
\centering
\includegraphics[scale=0.182]{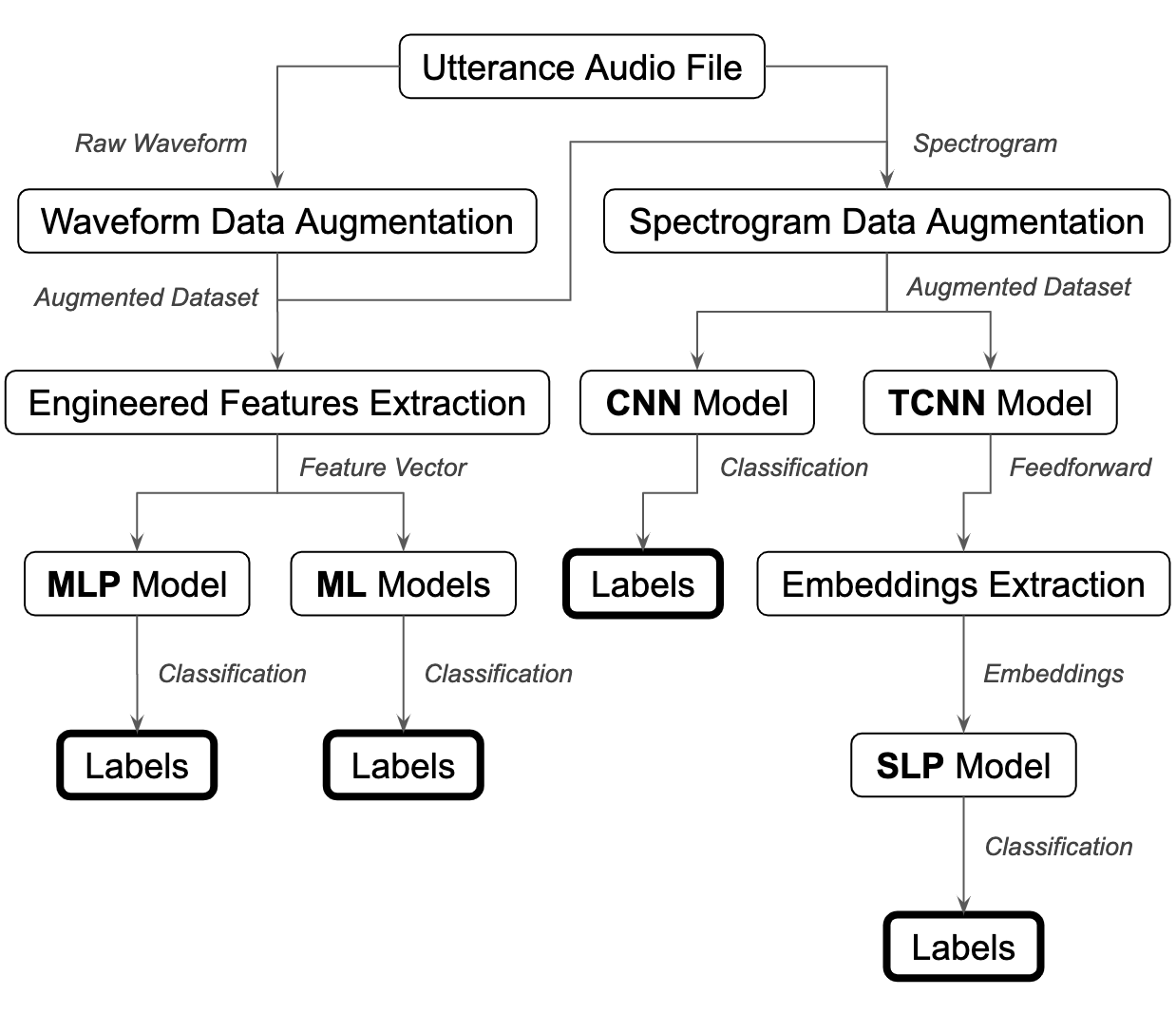}
\caption{Diagram illustrating the QVP workflow followed in this study. CNN stands for Convolutional Neural Network, TCNN for Triplet Convolutional Neural Network, MLP for Multi-Layer Perceptron, SLP for Single-Layer Perceptron and ML for other Machine Learning models.}
\end{figure}

\subsection{Dataset}
\label{s31}

The Amateur Vocal Percussion (AVP) \cite{34} dataset contains vocal percussion recordings from people with little or no experience in beatboxing. It includes 9780 utterances imitating four drum types (kick drum, snare drum, closed hi-hat, and opened hi-hat) recorded by 28 participants and with annotated onsets and labels. We use the personal subset, where the participants are asked to imitate the drums in their own way. For every participant, this subset is composed of five audio files: four files containing around twenty-five isolated utterances from the four classes (train set) and one file of arbitrary length with a beatbox-style improvisation using the chosen personal imitations (test set). The data belonging to the three discarded participants is included here for evaluation, as we also want to see how the models decide between labels when given ambiguous utterances that are hard to classify.

\subsection{Feature Extraction and Selection}
\label{s32}

Instead of feeding the raw waveforms directly to the machine learning models, a good practice in MIR is to compress the information into lower-dimensional feature vectors to make the analysis more manageable.

One of these arrays is the \textit{mel frequency spectrogram}, a time-frequency representation that compresses the frequency channels of the Fourier spectrogram into a small number of bands by applying a transformation based on the mel scale. We feed mel spectrograms from single utterances to the convolutional models in sections \ref{s342} and \ref{s343}. After fixing the length of all audio waveforms to 0.68 seconds (30k samples at 44100 hz), we calculate the mel spectrogram using a window size of 96 milliseconds and a hop size of $floor(1000\cdot\frac{30}{N_b})$, where $N_b$ is the number of mel bands (8, 12 or 16). By making the hop size dependent on the number of mel bands that way, we obtain square mel spectrogram representations (8x8, 12x12 or 16x16). This squareness property allows us to explore different hop sizes and frequency resolutions while making both data augmentation filtering and the design of network architectures more comfortable.

We also study another type of feature set composed of \textit{engineered features}. These are made out of audio low-level descriptors (e.g. spectral centroid) and statistical functionals (e.g. mean value), used to aggregate these descriptors through time. We feed these engineered features to the multi-layer perceptron in \ref{s341} and the rest of machine learning models in \ref{s344}. For each vocal percussion utterance, we compute 13 Mel Frequency Cepstral Coefficients (MFCCs), the zero-crossing rate, and several spectral descriptors (centroid, spread, skewness, kurtosis, flatness, and roll-off). We extract all of these using Librosa \cite{30}. In the case of statistical functionals, we compute the mean, standard deviation, minimum, maximum, and interquartile range of each descriptor through time. We have, therefore, 20 descriptors and 5 functionals, making a set of 100 engineered features.

Finally, before feeding these engineered features to the final classifiers, we use random forests to perform \textit{feature selection} in order to minimise overfitting and maximise accuracy. This way, random feature subsets are fed to decision trees that perform the utterance classification task in the usual way and then derive feature importances based on how many times they appear in these trees. We implement the random forest algorithm using the Scikit-Learn library \cite{32}. We also study the effect of selecting different amounts of features (1, 2, 4, 8, 16 and 32) to make the final set.

\subsection{Data Augmentation}
\label{s33}

The scarcity of training data is usually a matter of concern if this is high-dimensional. The \textit{curse of dimensionality} specifies that when our feature space is high-dimensional (e.g. waveform, spectrogram...) also is the solution space that the machine learning model is operating in. This causes sparsity in the data, making the model unable to obtain results with enough statistical reliability.

In our case, our training sets have only around one hundred utterances each, which is not an appropriate amount of data to train high-dimensional parametric models like neural networks. In order to tackle this problem, we use a regularisation technique known as \textit{data augmentation}, which applies realistic transformations to the original samples (e.g. a small amount of pitch-shifting) so to create new files and include them in the final dataset.

Two different data augmentation techniques are explored in this study: waveform data augmentation and spectrogram data augmentation.

\textit{Waveform Data Augmentation} is applied to the raw audio waveform of each file. We use three different methods: pitch-shifting (-1 and 1 semitones), as vocal percussion utterances sometimes include vowels with harmonic content; time-stretching (0.85 and 1.15), as utterances from the same class usually vary in length; and noise addition (0.005 and 0.01 amplitude factor from pink noise), which gently distorts the signal to let algorithms focus on the core timbral characteristics of the utterances. We use Librosa \cite{30} to make these transformations.

\textit{Spectrogram Data Augmentation}, which is based on the SpecAugment method for speech recognition \cite{31}, is applied to the spectrogram representation of each file and is composed by three different techniques: frequency channel zero-masking, time step zero-masking and time warping. With $N_b$ being the number of mel bands of the original spectrogram, we apply a random number of masks to each file in the range $[1, floor(2\cdot\frac{N_b}{8})]$, each of one of random width in the range $[1, ceil(\frac{N_b}{3})]$.

We experiment with serial combinations of these data augmentation techniques so to create expanded datasets that could improve performance accuracy for some models.

\subsection{Models}
\label{s34}

\subsubsection{Multi-Layer Perceptron}
\label{s341}

The Multi-Layer Perceptron (MLP) \cite{26} is the most simple feedforward neural network model, as well as the oldest. It has at least three layers of artificial neurons: an input layer with the extracted features, a hidden layer with learnt representations, and an output layer with the labels. These neurons are fully connected by trainable parameters called weights in a sequential way (input-hidden-output) and nonlinear activation functions are applied to neurons in the hidden and output layers, increasing the complexity of the learnt representation. Like most neural network models, MLPs use an optimization technique called gradient descent that is made possible using the backpropagation algorithm \cite{33}.

We feed engineered features to MLPs. The number of neurons in the hidden layers that we explore for these network architectures are [$\frac{n_i}{2}$], [$\frac{n_i}{2}$,$\frac{n_i}{4}$] and [$\frac{n_i}{4}$], where $n_i$ is the number of input (selected) features.

\subsubsection{End-To-End Convolutional Neural Network}
\label{s342}

Convolutional Neural Networks (CNN) \cite{23} are another type of feed-forward networks that include convolutional layers in their architecture. These convolutional layers are composed of several small local filter kernels that convolve with the input data, generating learnt feature maps as a result. The usual method to build end-to-end CNNs is to stack various convolutional layers to filter the data and create feature maps, pooling layers to subsample these maps, and fully connected layers (MLPs) to classify the samples.

We feed the two-dimensional mel spectrograms to different CNN architectures. The number of convolutional layers for these architectures are either 2, 4, or 6, and we also explore different number of filters for each of the layers (8, 16, 32, and 64). The kernel size of all convolutional filters is 3x3 with stride 1, while the kernel size corresponding to the max-pooling layers is 2x2 with stride 2, which downsamples the feature maps by a factor of 2. We experiment with different numbers of pooling layers (1, 2, and 3) and use batch normalization as an extra regularisation step.

\subsubsection{Triplet Convolutional Neural Network}
\label{s343}

A Triplet Convolutional Neural Network (TCNN) \cite{27} is a CNN that uses a special type of contrasting loss called the \textit{triplet loss} to make data embeddings (learnt features) that maximise separability between classes. It is, therefore, a learning-based feature extractor and not a classifier. This loss works by forming random data triplets with data samples: an anchor, a positive and a negative sample. The anchor and the positive samples are of the same class while the negative is not, and the goal of the algorithm is to minimise the distance (usually Euclidean) between the embeddings from the anchor and the positive samples while maximising the distance between the embeddings from the anchor and the negative samples. In summary, the algorithm tries to minimise the following function:

\begin{equation}
\label{eq1}
\mathcal{L}=\max (d(a, p)-d(a, n)+\textit{margin}, 0)
\end{equation}

, where $a$, $p$ and $n$ are the embeddings corresponding to the anchor, positive and negative samples respectively, $d(x,y)$ is a distance function applied to two arrays of embeddings and \textit{margin} is a fixed hyperparameter that regulates the target closeness and separation between the embeddings. We apply a Euclidean distance function and set the margin to 0 for the calculation of this loss.

After calculating the losses for all triplets in a training batch using equation \ref{eq1}, we need to make a final decision on which of these triplet losses we want to pass to the network for backpropagation \cite{28}. For instance, we may only want to pass the greatest loss value, the one belonging to the hardest triplet, or maybe pass the average value of all the triplet losses calculated for that specific batch. Two triplet selectors are explored for this purpose: random negative and hardest negative. We avoid averaging an elevated number of triplet losses, as this is known to cause instability in the training process \cite{28}.

Once the embeddings have been calculated, a Single-Layer Perceptron (SLP) classifier, composed of only an input and an output layer, is trained separately to extract the final performance accuracies. We build the networks with the same architectures used for the CNNs (section \ref{s342}) except for the output layer, which has 16 or 32 neurons (embeddings) instead of 4 (classes).

\subsubsection{Other Machine Learning Models}
\label{s344}

A final set of traditional machine learning methods is used for classification using engineered features. It includes the following algorithms: K-Nearest Neighbours, Support Vector Machine (with linear and radial basis function kernels), Decision Tree, Random Forest, AdaBoost, Naive Bayes, Quadratic Discriminant Analysis, and Single-Layer Perceptron. The values of the main hyperparameters that give each algorithm their best performance (e.g. number of tree estimators in the random forest) are found using grid search.

For this cluster of algorithms, the model that performs best is the one taken for evaluation in section \ref{s4}.

\subsection{Training Methodology}
\label{s35}

After standardising the input features, the models described in the last section are trained using different techniques. For the neural networks (MLP, CNN, TCNN), several hyperparameters and routines must be chosen. These parameters are the \textit{validation split}, which determines the percentage of samples used for network validation; the \textit{maximum number of epochs} that the network is allowed to reach during training; the \textit{optimiser type}, which makes the network gradually lower its performance error; the \textit{loss type}, which is the error metric used to guide the optimiser; the \textit{learning rate}, which dictates how fast the network is learning\footnote{If the learning rate is too high, it causes instability on the network's performance. If it is too low, it might take a long time to train the network.}; and the \textit{batch size}, which indicates the number of samples from the training set used to calculate a single loss value.

For all neural network models, we choose a validation split of 10\%, a maximum number of epochs of 200, and an Adam optimiser. We also investigate the effect of different learning rates on the training of the models, with eleven logarithmically spaced values in the range [0.1,0.000001]. We make sure that the chosen batch size is not larger than the tenth part of the training set. We assign a batch size of 8 to the original (non-augmented) datasets and experiment with sizes of 128, 512, and 1024 for augmented datasets. We choose cross-entropy as the loss function for both the MLP and the CNN (TCNN already includes the triplet loss). We choose all these parameters and parameter ranges in light of good practices for CNNs \cite{28}.

The networks also have two more sets of parameters that control the training process through the validation loss. These are the \textit{learning rate scheduler} and the \textit{early-stopping engine}. The former is concerned with reducing the learning rate by a factor $f_l$ for subsequent epochs if the validation error has not decreased in the last $N_l$ epochs. The latter automatically stops the training process if the validation error has not decreased in the last $N_e$ epochs. For all networks, we set $N_l$ to 0.1. For the MLP, we set $f_l$ to 2 and $N_e$ to 3; For the CNN, $f_l$ is 3 and $N_e$ is 8. Finally, for the TCNN, we have a $f_l$ of 2 and a $N_e$ of 3.

\section{Results}
\label{s4}

\subsection{Effectiveness}
\label{s41}

Final classification accuracies are gathered in table \ref{t1}. Random Forests (RF) with a hundred tree estimators were the best performing classifiers of all machine learning models in \ref{s344}. The values shown are the percentages of correctly classified utterances in the test set (improvisation) for all 28 participants.

We observe that neural network models performed best using the augmented set. They also show a preference for medium learning rates between 0.001 and 0.0001, which is in line with the observations in \cite{28}. Best results for CNNs are collected using spectrogram representations of sizes 12x12 and 16x16, while the best results for TCNNs are obtained with representations of size 8x8. Data augmentation seems to not have any major effect on the performance of machine learning models. Also, triplet models perform best when the loss from the hardest negative triplet (see \ref{s343}) is passed to the backpropagation algorithm.

\begin{table}[]
\label{t1}
\centering
\begin{tabular}{l|c|c|c|c|}
\cline{2-5}
& \multicolumn{1}{l|}{\textbf{MLP}} & \multicolumn{1}{l|}{\textbf{CNN}} & \multicolumn{1}{l|}{\textbf{TCNN}} & \multicolumn{1}{l|}{\textbf{RF}} \\ \hline
\multicolumn{1}{|c|}{\textbf{Original Dataset}} & 74.3 & 77.7 & 77.5 & 77.6 \\ \hline
\multicolumn{1}{|c|}{\textbf{Augmented Dataset}} & 76.9 & \textbf{82.2} & 80.9 & 78.3 \\ \hline
\end{tabular}
\caption{QVP classification accuracies (\%) of Multi-Layer Perceptron (MLP), Convolutional Neural Network (CNN), Triplet Convolutional Neural Network (TCNN) and Random Forests (RF).}
\end{table}

\subsection{Efficiency}
\label{s42}

We choose the algorithms that gave the best results for each of the models to evaluate performance efficiency (speed). In particular, we select the ones whose accuracies lie between the top accuracy and one percentage point below it. We include all preprocessing stages in the measurements (data augmentation, feature selection...). Training and testing times are averaged across the 28 participants, as model parameters and dataset lengths gave rise to very different training and testing times depending on the participant. All experiments are performed in the same machine (MacBook Pro 2015 laptop) and under the same conditions.

\begin{table}[]
\centering
\label{t2}
\begin{tabular}{c|c|c|c|c|}
\cline{2-5}
\multicolumn{1}{l|}{}& \textbf{MLP} & \textbf{CNN} & \textbf{TCNN} & \textbf{RF} \\ \hline
\multicolumn{1}{|c|}{\textbf{Training Speed}} & 14.08& 217.42& 28.83& 7.79\\ \hline
\multicolumn{1}{|c|}{\textbf{Testing Speed}}& 1.19& 1.46& 1.80& 1.44\\ \hline
\end{tabular}
\caption{Performace speeds for each type of model in seconds. Values are averaged across all participants.}
\end{table}

Results are shown in table \ref{t2}. Both the CNN and the TCNN models take augmented datasets as input, while the MLP and the random forest (RF) use the original dataset. We see that the CNN model, while being the most accurate one, it trains very slowly. Its testing speed, however, is comparable with those of the rest of the models. Hence, if waiting for training to finish is not an issue, the user of the QVP system could use an end-to-end CNN algorithm to classify utterances as fast as the other methods and with higher accuracy.

\subsection{Stability}
\label{s43}

In order to evaluate performance stability, we take a look at how each model performs when given different combinations of \textit{hyperparameter values}. This would be the first metric of performance stability. Then, once discovered the optimal hyperparameter values for each model (giving the accuracies in table \ref{t1}), we study how these specific algorithms perform when their internal parameters (e.g. decision tree's random state, neural networks' weights... etc.) are given by \textit{random initialisation}. This would be the second metric of performance stability.

The probability density functions of both measures of stability for all four models are shown in figure \ref{f1}. The shape of these distributions reveals whether the results of a particular model are strongly tied to the choice of hyperparameters or to the random initialisation of its internal parameters. In this way, we see that the most stable models to hyperparameter values are the two convolutional neural networks (CNN and TCNN), whose probability density functions showcase the least amount of statistical variance. When it comes to random initialisation of internal parameters, however, we see that random forests are more stable than neural networks. This could be explained by considering random forests as a type of ensemble learning (various decision trees deciding the most probable label), making them more prone to stability than the rest of the algorithms, which stand on their own.

\begin{figure}
\label{f1}
\centering
\includegraphics[scale=0.36]{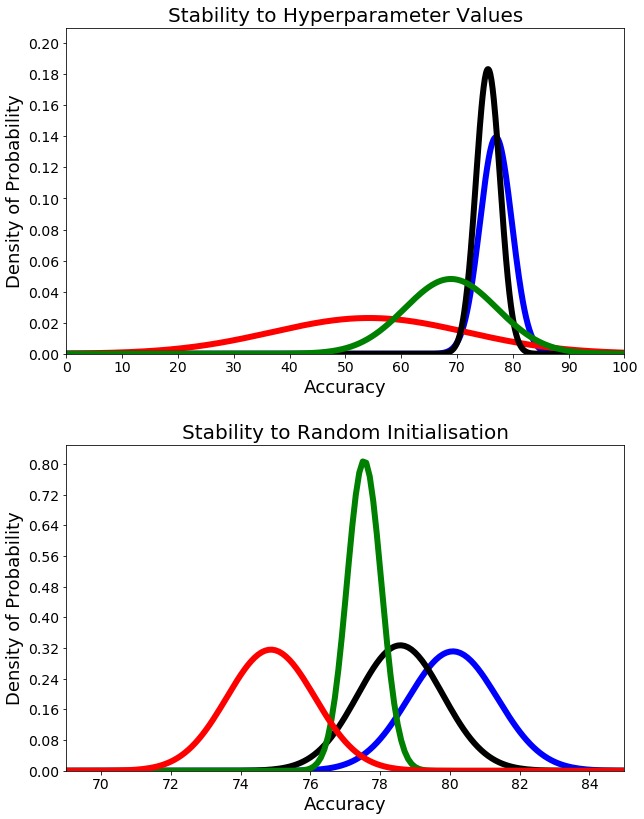}
\caption{Probability density functions to evaluate performance stability of the four models: MLP in red, CNN in blue, TCNN in black and RF in green.}
\end{figure}

\subsection{Interpretability}
\label{s44}

While it is challenging to interpret neural networks' decisions, models that take engineered features as inputs are generally more scrutable. For example, we can observe the kind of audio descriptors that were derived from the feature selection process (see \ref{s32}) so to then relate them to psychoacoustic phenomena that we can understand better. The most relevant selected descriptor across participants and models was the spectral centroid, which is often described as the ``centre of mass" of the frequency axis of the spectrum and happens to be one of the most important descriptors in sound timbre analysis. It was closely followed by the first MFCC and the third MFCC, which are also related to the utterance's low to high frequency ratio. For some participants, higher MFCCs were included in the top selected features, which encode more complex spectral information. They are possibly used to separate vocal imitations of closed hi-hat and opened hi-hat, which are often close in timbral similarity.

Related to the last point, the most common mistake for all four models is the confusion between closed hi-hat and opened hi-hat vocalisations. Many participants imitate them in a similar manner and, while the utterances are carefully pronounced when recording the isolated audio files for training, the diction within the improvisation files is sometimes erratic, often mixing both sounds in a hybrid utterance that the algorithm misclassifies in the testing stage. A greater pre-emphasis in the higher frequency bands for feature extraction or classification could improve accuracies for many participants.

\section{Conclusions}
\label{s5}

We have explored here several approaches for effective and efficient Query By Vocal Percussion. Different machine learning models and data augmentation routines have been tested, and several combinations of their hyperparameters have been investigated and later presented in the results section. We have compared the approaches in terms of effectiveness, efficiency, stability and interpretability. We found that the combination of data augmentation and convolutional neural networks achieved the best classification accuracies in our setting, at the expense of significantly longer training time and lower performance stability to random initialisation. In the future, we intend to carry out experiments in high-resolution percussive onset detection and build a quick and accurate user-based QVP system with our results here in mind.

\begin{acknowledgments}

$
\begin{array}{l}
\includegraphics[scale=0.1]{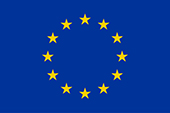}
\end{array}
$ This project has received funding from the European Union's Horizon 2020 research and innovation programme under the Marie Sk$\l{}$odowska-Curie grant agreement No. 765068

\end{acknowledgments} 

%%%%%%%%%%%%%%%%%%%%%%%%%%%%%%%%%%%%%%%%%%%%%%%%%%%%%%%%%%%%%%%%%%%%%%%%%%%%%
%bibliography here
\bibliography{bibliography}

\end{document}